\documentstyle[referee,psfig]{mn_latex}

\begin{document}

\title{On the use of Principal Component Analysis
in analysing Cepheid light curves}
%{Extra Stuff}
\author[S. M. Kanbur et al.]
{S. M. Kanbur D. Iono, $^1$\thanks{email: shashi@astro.umass.edu}
 N. R. Tanvir,$^2$ M. A. Hendry,$^3$\\ 
$^1$Department of Astronomy, University of Massachusetts, Amherst, MA 01003\\
$^2$Department of Physical Science, University of
Hertfordshire, College Lane, Hatfield, AL10 9AB, UK \\
$^3$Department of Physics and Astronomy, University of Glasgow,
Glasgow, UK}
\maketitle
\begin{abstract}
We show how Principal Component Analysis can be used
to analyse the structure of Cepheid light curves. This
method is more efficient than Fourier analysis
at bringing out changes in light curve shape as a 
function of period. Using this technique,
we study the shape of fundamental and first overtone
mode Cepheid light curves in the
Galaxy, LMC and SMC over a wide
period range. For fundamentals, we find evidence for structural changes at
$\log P \approx 1.55, 2.1$. It is suggested that the feature
at $\log P \approx 2.1$ is associated with a resonance in the
Cepheid normal mode spectrum. For overtones, we recover the
Z shape in the $R_{21}$ period plane and reproduce the
metallicity dependence of this Z shape.
\end{abstract}


\section{Introduction}

Cepheids play a vital role in astrophysics. Their pulsational
properties pose constraints that stellar pulsation and evolution
theories must satisfy simultaneously. Further they are the keystone
of the extra-galactic distance scale through the period -- luminosity
(PL) relation. In this paper, we describe a new way
of analysing the structure of variable star light and
velocity curves, Principal Component Analysis (PCA).
The technique is more general and more
efficient at describing structure than Fourier
analysis (Hendry et al 1999, Kanbur et al 2000a), Tanvir et al 2000).
Hendry et al 1999 also analyzed the period-luminosity-light curve shape relation
for Cepheids in the Galaxy and LMC. This was extended to the SMC by
Ligeza and Schwarzenberg-Czerny (2000).
In this paper we use PCA to examine light curve structure to search for other
resonances which have been claimed in the literature (Antonello 1998). 
We identify two features in the light curve structure of
long period Cepheids at $\log P \approx 1.55, 2.1$ days.
The feature at $\log P \approx 2.1$ 
may be connected with resonances in the Cepheid normal mode
spectrum. The templates generated with PCA have also been
used, together with Fourier decomposition, to fit sparse and noisy data to obtain good
period and magnitude estimates (Tanvir et al 1999, Kanbur et al 2001).

Section 2 describes Fourier analysis as applied to variable star data and
section 3 summarizes PCA. Sections 4 and 5 present our
results using this method. Section 6 presents our conclusions
and discussion and suggestions
for further work. 
\section {Previous Work}

The technique of Fourier decomposition has been used for some
time to analyse variable star data (Schaltenbrand and
Tammann, 1971). In the seventies the method was revived by Simon and
Lee (1981) who fitted expressions of the form,
$$A_0 + \sum_{k=1}^{k=N}(A_kcos(k\omega t + {\phi}_k))\eqno(1)$$
to observed data. They plotted the relative Fourier parameters,
$$R_{k1} = A_k/A_1, {\phi}_{k1} = {\phi}_k - k{\phi}_1 \eqno(2)$$
describing light curve structure
against period. These authors noted sharp breaks in the progression of
${\phi}_{k1}$ and $R_{k1}$ against period at a period of 10 days. This was
associated with the Hertzsprung progression, where a bump on the
descending branch of short period ($P < 10$ days) Cepheids moves
to the ascending branch of long period ($P > 10$ days) Cepheids. Using
linear adiabatic models, Simon and Schmidt (1976) and Simon
and Lee (1981) interpreted the
Hertzsprung progression and the sharp break in the Fourier parameter plots
at 10 days as evidence of a resonance between the fundamental mode,
$P_0$, and the
second overtone, $P_2$, such that $P_2/P_0 = 0.5$ at $P_0 = 10$ days.
Since then, Fourier decomposition has been used to identify possible
resonances in first and
second overtone Cepheids (Antonello and Aikawa 1995, Antonello and
Kanbur 1998). Microlensing surveys (MACHO, OGLE and EROS)
have also made significant use of Fourier analysis (Welch et al. 1996,
Beaulieu and Sasselov 1996, Udalski et al. 1999). Recently, Antonello and
Morelli (1996) and Antonello (1998) suggested the presence of
resonances in the normal
mode spectrum of long period Cepheids,
namely $P_0/P_1 = 3/2$ at $ 1.34 < \log P_0 < 1.40$ days,
$P_0/P_3 = 3$ at $1.40 < \log P_0 < 1.43$ days and
$P_1/P_0 = 0.5$ at $1.95 < \log P_0 < 2.13 $ days.
Their argument was based on a Fourier
analysis of observed light and velocity curves. Figure 1
is taken from Antonello and Morelli (1996) and shows a plot of the
Fourier parameters $R_{21}$ and ${\phi}_{21}$ plotted against period. Though it
may be argued that there are breaks in this plot at periods around
20-25 ($\log P =1.3$ to $\log P = 1.4$)
days (see also figures 1 and 2 of Antonello and
Morelli 1996), definitive conclusions
are hard to draw. Figure 2 of Antonello (1998) extends the period
range to about 150 days and was used by Antonello and co-workers to suggest the
presence of a resonance at a period between 90 and 134 days.
 
\section{Principal Component Analysis}

Kanbur et al (2000a)), Hendry et al 1999 suggested the
possibility of using the technique of
Principal Component Analysis (PCA) in studying variable star data.
Here we 
review and extend the formalism briefly described in Kanbur et al (2000a)). Let
$X_{ij}$ be the $j^{th}$ ($1 \le j \le P$) observed
point on the $i^{th}$ ($1 \le i \le N$) light curve. The input matrix
of the data is
$$S_{jk} = {1\over{N}}\sum_{i=1}^{i=N}X_{ji}X_{ki},\eqno(3)$$
sometimes referred to as the covariance-variance matrix
(Murtagh and Heck 1987).
This measures the relationship between the $j^{th}$ and $k^{th}$
points, averaged over all $N$ stars in the sample. In our
use of PCA, we seek to
write any Cepheid light curve as a linear combination of elementary
light curves, $u_i^t$,
$$V(i) = \sum_{t=1}^{t=P}PC_t(i)u_i^t.\eqno(4)$$
Here, $PC_t(i)$ are Principal Component coefficients 
so that $PC1$ for star $i$ is $PC_1(1)$ etc.
One example of such elementary light curves
are the harmonics in equation (1). Given the
assumption that the elementary light curves are harmonics, then
fitting a Fourier expansion to variable star data will find
the best coefficients $PC_t(i)$ in a least squares sense. That is,
a Fourier fit will yield estimates for $A_k$ and ${\phi}_k$ in
equation (1) that minimize the sum of
squared deviations in magnitude
between the model and the observed data. 
The technique of PCA will optimise,
in a least squares sense, the fit of the data to the model
of equation (4) without any prior assumption about the
nature of the functions $u^t_i$ (Murtagh and Heck 1987);
instead the $u^t_i$ and their coefficients are
determined entirely by the properties
of the input matrix, $S$. 
Further the functions
${u_i^t}$ are orthogonal to each other and hence the elementary
light curves in (4) are distinct from each other.
It can be
shown that this optimal set of curves, 
$\{u^t\}$, is given by the solution of the eigenvalue equation,
$$Su = {\lambda}u.\eqno(4) $$
Solution of this equation yields ${u_i^t}$ and a ${\lambda}^t$ for
each vector $u_i^t$. After suitable normalization, the
${\lambda}^t$ can be interpreted as the percentage variance in the
light curve data explained by the $t^{th}$ light curve.
We can project each light curve onto the eigenvectors $\{u_i^t\}, t=1,...,P$,
by forming the sum,
$$PC_t(i) = \sum_{j=1}^{j=P}X_{ji}u^t_j.\eqno(5)$$
Each Cepheid light curve $V(i), i = 1,..,N$ can then be represented as
$$ V(i) = \sum_{t=1}^{t=P}PC_t(i)u_i^t.\eqno(6)$$

In this work, we convert the Fourier expansion given in equation (1) to the
equivalent form,
$$A_0 + \sum_{k=1}^{k=N}(a_kcos(k\omega t) + b_ksin(k\omega t)),\eqno(7)$$
where,
$$A_k^2 = a_k^2 + b_k^2, tan({\phi}_k) = b_k/a_k,\eqno(8)$$
so that the X matrix consists of
$$X_{ij} = a_{ij},X_{ij+1}=b_{ij}.$$
Therefore the resulting eigenvectors obtained from solving equation (4), that is the optimal
basis set, are in fact a set of vectors consisting of $a,b$ coefficients so that equation (6)
becomes,
$$V(i) = \sum_{j=1}^{j=P}PC_j(i)(\sum_{k=1}^{k=P}(x_k^jcos(k\omega t) + y_k^jsin(k\omega t))).\eqno(9)$$
In this equation, the $x_k^j,y_k^j$ are fixed and obtained from the PCA analysis by the solution of
equation (4). We can find a relation between Fourier coefficients and PCA coefficients,
$PC_j(i)$, by equating $cos(k\omega t), sin(k\omega t)$ coefficients between equation (9) and (1).
This yields,
$$A_k^2 = {\big(\sum_{j=1}^{j=P}PC_jx_j^k\big)}^2 + {(\sum_{j=1}^{j=P}PC_jy_j^k)}^2.\eqno(10)$$
$$tan({\phi}_k) = - {{\sum_{j=1}^{j=P}(PC_jy_k^j)}\over{{\sum_{j=1^j=P}(PC_jx_k^j)}}}.\eqno(11)$$
This establishes a {\it direct} correspondence between the PC coefficients and the Fourier parameters
which are plotted against period. In particular at the 10 day resonance, $R_{21} = A_2/A_1$ goes down.
Plotting $A_1, A_2$ against period it is easy to see that this is because $A_2$ goes down at 10
days. If $A_2$ goes down then equation (10) implies that either $PC_1$ goes down or $PC_2$
goes down or both go down. This proves that if there is a change in the structure of the light curve
as shown by a change in the Fourier parameters, equations (10) and (11) guarantee that such a change
will be reflected in the PC coefficients. It would be interesting to apply the amplitude
equation formalism developed by Buchler (1993) to understand the physical nature of the PCA
coefficients perhaps through the use of equations (10) and (11). Our
initial work suggests that the $PC1$ coefficient is correlated with
amplitude. We intend to pursue this, coupled with a
PCA study of hydrodynamic model light and velocity curves in future work.

In a practical
sense, the principal advantage
of PCA is efficiency. PCA requires 4 to 6 parameters to
describe Cepheid light curve structure (including bump Cepheids, Kanbur et
al 2000a)) whereas an $8^{th}$ Fourier fit needs 16 parameters. It is true that $A_1, R_{21}$
and ${\phi}_{21}$ - that is three Fourier parameters are often used in comparing models
and observations but in order to get stable values of these coefficients, a 6 to 8 order
Fourier fit is needed.

We can plot the $PC_t(i)$ against period analogously to the
way Fourier parameters in equation (2) are plotted against 
period. Recall that figure 1 shows a plot
of the Fourier parameters $R_{21}$ and ${\phi}_{21}$ plotted against
period for the Cepheid data presented in Antonello and Morelli (1996). The upper
panel, which shows ${\phi}_{21}$ against log period, indicates clearly
a discontinuity at $\log P = 1$ but little structure thereafter.
The lower panel, which shows $R_{21}$ against log period also
exhibits clear structure at $\log P \approx 1$, followed by
a general rise to $\log P \approx 1.4$ and some scatter thereafter. It should be
noted that Antonello (1998) included more data on longer period Cepheids and found
some evidence of a decrease at large ($\log P \approx 2$) in the $R_{21}$ - period
plane. 

For our PCA analyses,
we used V band data
from the Galaxy, LMC and SMC published by Moffett and Barnes (1989) and
Berdnikov and Taylor (1995), and Antonello (1998) to
carry out such a procedure. We emphasize that all
this data has {\it good} Fourier decomposition such that the
light curve obtained from the Fourier decomposition is
an excellent representation of the actual data points with
few numerical bumps or wiggles.
The top and bottom panel of
figure 2 shows
plots of $PC_1(i)$ and $PC_2(i)$ against the logarithm of the
period (see equation 5).
Open circles denote
Galactic Cepheids, solid circles and open stars are the LMC and SMC
respectively. The data used were taken from the McMaster web site.
Typical error bars (see section 4) are shown in the top left hand corner of the upper
panel. This error bar is applicable to all subsequent plots of the PC
coefficients.
Taken together, these first two principal components account for $97$ percent
of the variation in the data.

The top panel, showing 
the first principal component, PC1, plotted against
the logarithm of the period displays some scatter for $\log P$
less than one, an  abrupt decrease at $\log P \approx 1$ and then a rise from $\log P$ = 1 to $\log P$ = 1.2. There is
a "plateau" till about $\log P = 1.55$, after which PC1 falls gradually till
$\log P = 2.1$. The bottom panel, which shows the second principal component,
PC2, shows a similar pattern to the top panel, but with significantly smaller
scatter. In particular the break and sharp decrease at $\log P = 1$ is very clean, the general
rise from $\log P = 1$ to $\log P = 1.55$ has
little scatter. There is a sharp turnover at $\log P = 1.55$ and
a decline to a minimum at $\log P = 2.1$. We notice that PC1 is always
positive while PC2 can be either positive or negative. This is because
each light curve is written as equation (6), with the first term in this sum,
$PC_1(i).u_1(i)$, being the basic light curve, while the second term, $PC_2(i).u_2(i)$
is a second order term showing corrections to the basic light curve. It
can be argued that features like bumps are second order effects and should be
more evident in the second and higher principal components.

Equation (10) and the shape of the plots in figure 2 enables us to
categorically state that we see the presence of the resonance at $\log P = 1$
in the PC plots.
The bottom panel shows a sharp drop
at $\log P = 1$ followed by a gradual
rise in PC2 from $\log P = 1$ to $\log P = 1.55$ and a gradual
decline thereafter till $\log P = 2.1$. In both panels, we see 
three stars with PC1, PC2 values significantly
lower than normal at $\log P \approx 1.6-1.7$, and one
star with a significantly higher than normal
value at $\log P \approx 1.8$. We do not treat these stars here
but note that our
error analysis, described in the next section,
suggests that the uncertainties on the PC1/PC2 values are small.
Despite this, figure 2 clearly shows a definite maximum
at $\log P = 1.55$ and a well defined decrease thereafter
till $\log P = 2.1$. This latter minimum is
dependent on 4 stars. The data for these stars is
exactly as used by Antonello and Morelli (1996), and
Antonello (1998) with sufficient phase coverage to permit an excellent
Fourier decomposition. We suggest that this
rise and fall in the PC1/PC2 values is real and caused by pulsation physics.
We
believe that figure 2 and constitutes convincing evidence
that the change in the structure of light curve shape
at $\log P = 2.1$ is very similar to
that at $\log P = 1$.
Since a resonance is
associated with the latter, it is our contention that the
feature at $\log P = 2.1$ is associated with a resonance in the Cepheid
normal mode spectrum. This has already been
suggested by Antonello and Morelli (1996) using
Fourier decomposition as in figure 1 as evidence.  We feel that our figure 2
comprises a much
stronger case than has hitherto been presented.
One caveat with this is that changes in light
curve shape are not necessarily associated with resonances (Kienzle et al,
1999). 

\section{Error Analysis}

PCA has the nice property that if the data are normally distributed, then
$${\sigma}^2(PC_t(i)) = \sum_{j=1}^{j=P}({\sigma}^2(X_{ij})u^t_j.\eqno(6)$$
$X_{ij}$ is the $j^{th}$ observed point on the $i^{th}$ star.
However, with the
specification that phase 0 is maximum
light, we have chosen to parameterize the light curves by the Fourier
expansion,
$$A_0 + \sum_{k=1}^{k=P}[a_kcos(k\omega t) + b_ksin(k\omega t)],\eqno(7)$$
which is entirely equivalent to equation (1). Now our X matrix
consists of
$$X_{ij} = a_{ij}, X_{ij+1} = b_{ij},$$
where the $a_{ij},b_{ij}$ are the $a_j,b_j$ in expression (7) for star i.
This can be viewed as an initial guess. The PCA approach uses this to find the
"optimal" way to represent light curves via equation (4). Consequently,
the quality of the Fourier decomposition used has a bearing on the
conclusions of this paper. As stated previously, the Fourier decompositions
used were those carried out by the original authors of the data used.
Superimposing the light curve produced by the Fourier decomposition
on the original data points yielded an excellent match with no numerical
bumps or wiggles in the Fourier decomposition light curve. Hence
in our application of PCA, we require good enough phase coverage
to permit a reasonable Fourier decomposition. Given this
situation, it is our thesis that PCA is more efficient at
bringing out significant changes in light curve shape.

In obtaining the fit given in (7) to the actual data, we can write the 
problem as
$$\underline{V} = \underline{A}\underline{\beta},$$
where $\underline{V}$ is the vector of data, the observed magnitudes,
$\underline{A}$ is a known matrix and $\underline{\beta}$ is the
vector of unknowns, consisting of the $a_k, b_k$ in equation (7).
A least squares estimate $\hat{\beta}$, for $\beta$ can then be obtained.
If the photometric errors are normally distributed, as is normally the
case,
$$\underline{V} = \underline{A}\underline{\beta} + \underline{\epsilon}$$
where,
$$\epsilon \sim N(\underline{V},{\sigma}^2),$$
then it can be shown that $\hat{\beta}$ is normally distributed. Since
the $\underline{\hat{\beta}}$ consist of the $a_{ij},b_{ij}$, which
make up the input matrix to the PCA, our input data is normally
distributed and hence equation (6) applies. Moreover, ${\sigma}^2(X_{ij})$ 
is obtained easily via standard numerical
techniques like Singular Value Decomposition. This enables a complete
estimate of the error on $PC_t(i)$ which is the quantity considered in this
paper. If the phase coverage is good, then the formulae given by
Petersen (1986) can be used as an accurate estimate of ${\sigma}^2(X_{ij})$.

Figure 3 shows the bottom panel of figure 2 but with error estimates derived using the
Petersen (1986) formulae. The larger error
bars near $\log P = 1.5$ are due to a poorer Fourier fit since bad fits
are reflected
in the large standard deviation of the Fourier
fit and the subsequent large error in the Principal Component diagrams.
It can also be the case that numerical wiggles are present in the
Fourier fit despite a small standard deviation. This was not the case with
the data used in this study. In any case, such "wiggles" would be contained
in the higher order Fourier parameters and hence in the higher order
principal components.
Nevertheless the figure
clearly shows that the structure of 
figure 2 is real. Further these error bars are not significantly
altered if photometric errors associated with the data are included
in the error analysis.

\section{Asymmetry parameter}

Antonello (1993) used the asymmetry parameter, defined
as $(M-m)$, where $M$ and $m$ are the phase of
maximum and minimum light respectively.
Using the same data as shown figure 1,
we plot $1-(M-m)$ against log period
in figure 4.
Again we
see that at 10 days, a sharp change occurs in this asymmetry parameter.
On either side of the resonance, minimum light occurs at a phase of
about 0.7, while at the resonance it occurs at a phase of 0.5. The
light curve shape near the resonance becomes more sinusoidal, departing from
the usual saw tooth shape.
While this behavior deserves further investigation, in
the context of the present paper, we note that there is a similar change in the
asymmetry parameter at $\log P \approx 2.1$.
The feature at
$\log P \approx 1.55$ is the reverse of what happens at a resonance -
the phase of
minimum light occurs even later than usual and the light curves take on a
more saw tooth like appearance. We comment on this later and suggest figure
3 as further evidence of a resonance at $\log P = 2.1$.

\section{I band data}

We performed the PCA analysis on I band data Cepheid data again from
Moffett and Barnes (1989), and Berdnikov and Turner (1995). Note that
this was the same data as for our V band analysis without the extra
long period stars studied in Antonello (1998). Figure 5 shows
the first two principal components from the I band data. They are generally
similar to the V band figures. The points form a clump for stars with
periods smaller that $\log P = 1$. At $\log P = 1$, there is a sharp
discontinuity to negative values of PC1/PC2. From $\log P = 1$, PC1
gradually rises to a peak at about $\log P = 1.6$. After this PC1/PC2
fall again to negative values at $\log P = 2.1$. Since there are
theoretically expected differences in a Cepheids light curve as
the wavelength changes, we would expect some differences but the fact that
the general structure is the same indicates strongly that PCA is
a bona fide technique in analysing variable star light curve structure.

In this study, we performed the PCA of V and I band data
separately but there is no reason why the V and I band data cannot be
analysed jointly. In fact such an approach could lead to
insights into structural changes across period and wavelength.
We leave a detailed comparison of the V/I band PCA analysis to a future paper.

\section{EROS data}

EROS (Beaulieu et al. 1995,)
is a microlensing search that has resulted in Cepheid light
curves for some 550 fundamental and first overtone
Cepheids in the LMC and SMC. Figure 6 and 7 shows plots of the first two
principal components plotted against period. These data were kindly provided by
Beaulieu (2000, private communication). The open and dark circles
represent fundamental and first overtone oscillators respectively.
We see a clear differentiation between the two types in both
components.
Figures 8 and 9 show the first two principal components plotted against
period for fundamental mode Cepheids from the SMC and LMC. Open and
dark circles represent the LMC and SMC respectively. We see clearly
the position of the resonance at $\log P = 1$, confirming with another data
set that PCA can pick up the presence of this resonance. We note
that at given period, there is considerable scatter.
Using our error analysis techniques, it can be
shown that the error bars on these points are considerably less
than the scatter at a given period. Hence this
scatter is real and presumably caused by
differences in global stellar parameters leading to
differences in light curve structure. The SMC stars extend to shorter
period and there is a tendency for shorter period stars to have 
higher values of PC2/PC1.

\section{Overtones}

It is well known that first overtone Cepheids show a change in the shape of the first overtone light curve at 
3.2, 2.7 and 2.2 days for the Galaxy, LMC and SMC
respectively (Beaulieu, private communication). To investigate how the PCA technique works in this regard, we
present in figures 10-12 plots of the PC coefficients for Galactic, LMC and SMC first
overtone stars. Figure 10 uses Galactic overtone data taken from Antonello and Poretti (1986).
No discernible pattern is visible here, perhaps due to the relatively few data
points. Figure 11 shows the the first two PC
components for the EROS data with the LMC and SMC represented by
open and solid circles respectively. There is significant
structure but no definitive conclusions can be made about changes in
structure at periods close to 3.2 and 2.7 days. Because of this, we also analyzed the OGLE survey data (Udalski et al 1999) which has detected about
500 and 800 overtones in the LMC and SMC respectively.
Figure 12 shows the first two principal
components for the LMC and SMC using OGLE first overtone data. If we initially
look at the PC2 plot, for periods greater than about $\log P \approx 0$, we
see an increase in PC2 with period to a maximum followed by a decline.
The maximum for the LMC and SMC is at $\log P \approx 0.5$ and $\log P 
\approx 0.4$ respectively. These periods correspond closely to the periods at
which changes in the structure of plots of $R_{21}$ against period are seen
for overtone light curves in the LMC and SMC (Udalski et al 1999). There is considerable
structure in figure 12 which will be treated in a future paper but the main
point for the purposes of this study is that the PCA approach picks out
changes in the structure of overtone light curves as a function of
metallicity.

\section{Discussion}

We have developed a new way of analysing the structure of 
Cepheid light curves, Principal Component Analysis, which is
much more efficient at bringing out changes in light curve structure than
Fourier analysis. The method is particularly suitable to analyse
the vast quantity of light curves produced by the MACHO, EROS and OGLE projects.
We emphasize that our comparison between PCA and Fourier analysis is made only
for those situations when a reliable Fourier decomposition already exists.
In this case, PCA can be used
to reproduce the light curve with about
half the parameters needed by the
Fourier decomposition technique. 
We have shown that PCA is a powerful method with which to detect changes in
light curve shape.
Since light curve shape changes can be associated with the resonances, PCA will
be a powerful tool to search for resonances in observed Cepheid
light {\it and} velocity curves since there is no
reason why this technique cannot be applied to velocity
curves.
The reliable
detection of 
resonances is important because this can be used to
place important constraints on global stellar parameters such as the mass,
luminosity, effective temperature and metallicity (Simon and Kanbur 1994). 
 
Specifically, we have found that based on
a PCA analysis of Cepheid light curves and an examination of
the asymmetry parameter, the structure of Cepheid light curves
exhibit similar features at $\log P = 1$ and $\log P = 2.1$ in the following sense:
at both periods, the phase at minimum light is close to 0.5 and
the $PC1$, $PC2$ coefficients attain local minima. The approach to these local
minima is different in the two cases but nevertheless there is a local minima at
both periods. This clearly suggests a significant change in the shape of the
light curve at these two periods. Since equations (10 and 11) establish a direct link
between Fourier parameters and PC coefficients, this also implies a change in
the Fourier parameters at these periods. Indeed figure 2 of Antonello (1988)
has shown some evidence that the ratio $R_{21}$ goes down at $\log P \approx 2.1$.
We suggest that our results strengthen the case for an important change in the light curve
shape at $\log P \approx 2.1$.
Model calculations are needed to confirm if a resonance is indeed present and is
associated with the fundamental and first overtones $P_1/P_0 = 0.5$ as suggested by Antonello (1988).
Another feature present in the PCA
and asymmetry plots is the 
maximum at $\log P = 1.55$. At this
period, the light curves become less symmetric and the PC2 coefficient
reaches its largest value. We note that Udalski et al (1999) show
a plot of $R_{21}$ against period for SMC Cepheids which also
suggests a similar
break at $\log P = 1.55$ though the number of stars
with periods longer than $\log P = 1.55$ in their
sample is too
small to be definitive. Based on our PC2 plots, we
conclude that this change in light curve shape is
real and plan to investigate its cause. One
possible way to sharpen this is to study light curve
structure for long period ($\log P > 1$) Cepheids and examine the
spread at given period as a function of other stellar parameters such
as mean color.
These topics will be the subject of a future paper.  

Obviously more short period ($\log P < 1.7$)
have been observed than long period ($\log P > 1.7$) Cepheids.
This is caused both by the fact that it is difficult to
obtain adequate phase coverage, sufficient for Fourier
decomposition, for such long period stars, and possibly also
because such long period stars may be intrinsically rare. However, the
possible rarity of such stars does not affect the suggestion that a
resonance occurs
in the normal mode Cepheid spectrum at long periods. Since Cepheids in low
metallicity environments will have higher luminosities and hence
longer periods for the same mass, the best chance to
observe such stars is in low metallicity galaxies like the
SMC. In fact the majority of the long period Cepheids used in
this study were from the LMC/SMC.

Antonello and Morelli (1996) also suggested resonances in the period
range $1.38 < \log P < 1.43$. Our analysis of the first two principal
components (PC1/PC2) does not show any features in this period range.
It is possible that these features may be visible in the higher order 
principal components and we leave this for future work.   
We also intend to apply PCA to the analysis of
first overtone light and velocity
curves. The greater efficiency of the method described here should
enable a decisive contribution to answering where, if any, the 
resonance occurs ($P_4/P_1 = 0.5$ at $P_1 = 3.2$ days (Antonello
and Poretti (1986), or $P_4/P_1 = 0.5$ at $P_1 = 4.58$ days
(Kienzle et al 1999).

\section{Acknowledgements}

We thank the referee, J.P> Beaulieu for helpful suggestions and comments.

\begin{figure}
%\vspace{9cm}
\centerline{\psfig{file=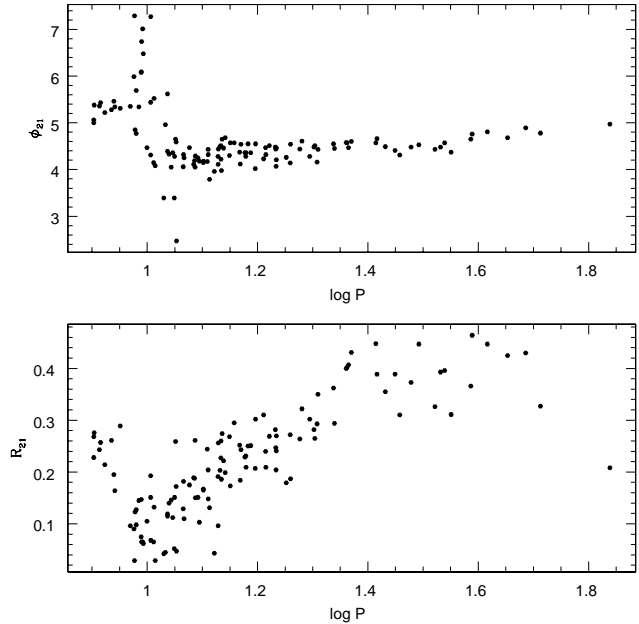,height=9cm,width=9cm}}
\caption{Plot of Fourier parameters against Period for
long period Cepheids.}
\end{figure}

%\begin{figure}
%\centerline{\psfig{file=pca1.ps,height=9cm,width=9cm}}
%\caption{The first Principal Component against Period. Open circles,
%galaxy, solid circles, LMC, open stars, SMC.}
%\end{figure}

%\begin{figure}
%\centerline{\psfig{file=pca2.ps,height=9cm,width=9cm}}
%\caption{The second Principal Component against Period. Open circles,
%galaxy, solid circles, LMC, open stars, SMC.}
%\end{figure}

\begin{figure}
\centerline{\psfig{file=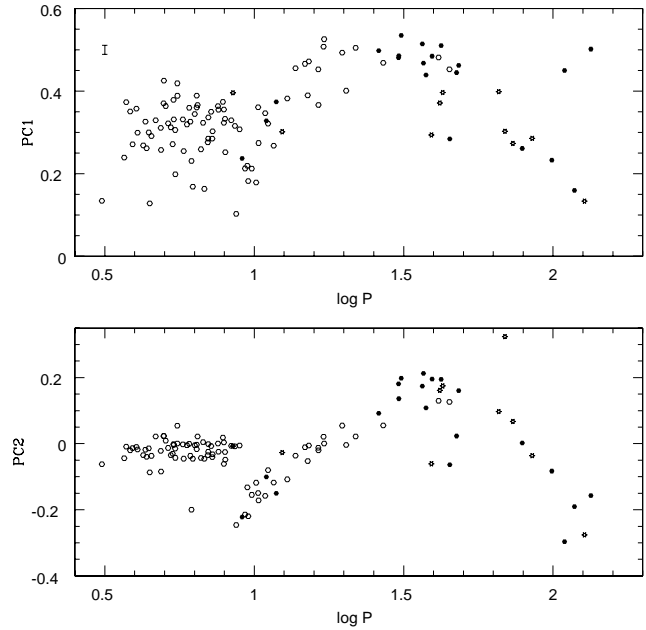,height=9cm,width=9cm}}
\caption{The first two Principal Components against log period.
Open circles, galaxy, solid circles, LMC, open stars, SMC. Typical error
bar plotted in  top left hand side.}
\end{figure}

\begin{figure}
\centerline{\psfig{file=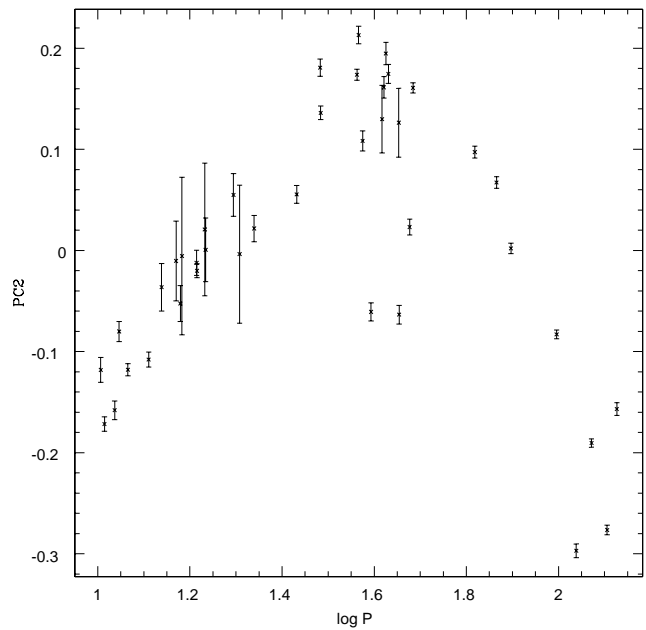,height=9cm,width=9cm}}
\caption{Errors associated with the second Principal Component plot}
\end{figure}

%\begin{figure}
%\centerline{\psfig{file=fdplot.ps,height=9cm,width=9cm}}
%\caption{Fourier fitted light curve of the star with the largest error bar in
%figure 4}
%\end{figure}

\begin{figure}
\centerline{\psfig{file=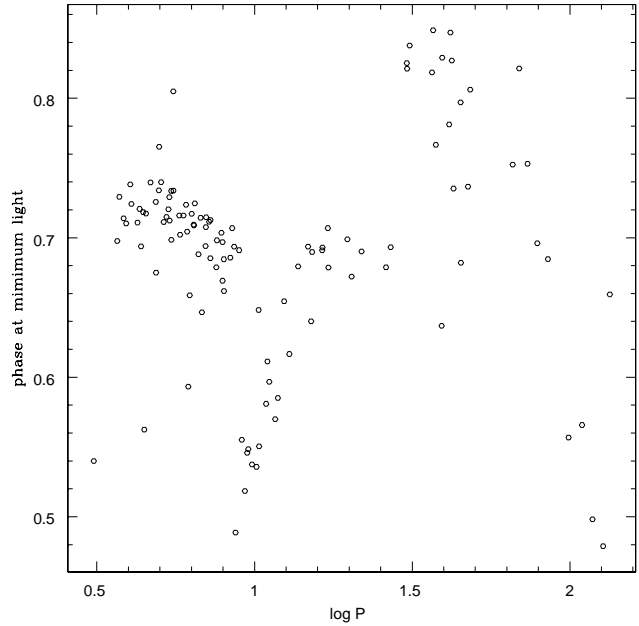,height=9cm,width=9cm}}
\caption{Phase of Minimum Light against Period}
\end{figure}

\begin{figure}
\centerline{\psfig{file=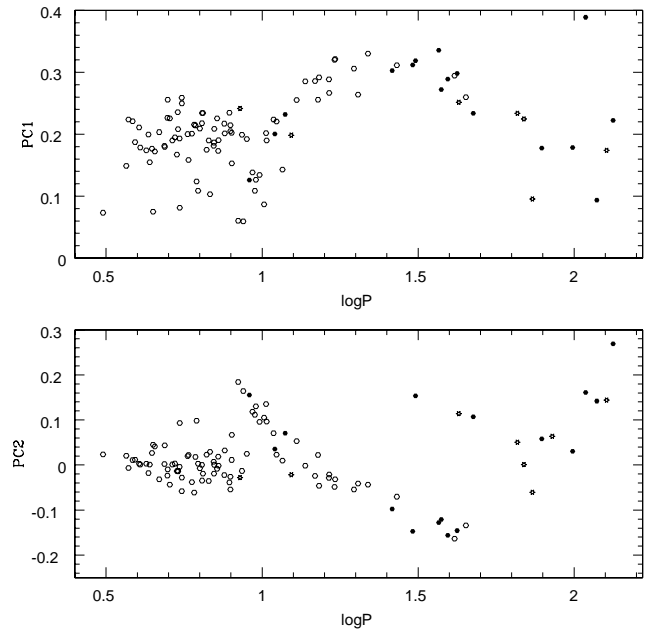,height=9cm,width=9cm}}
\caption{First two Principal Components plotted against period
for I band Cepheid light curves}
\end{figure}

\begin{figure}
\centerline{\psfig{file=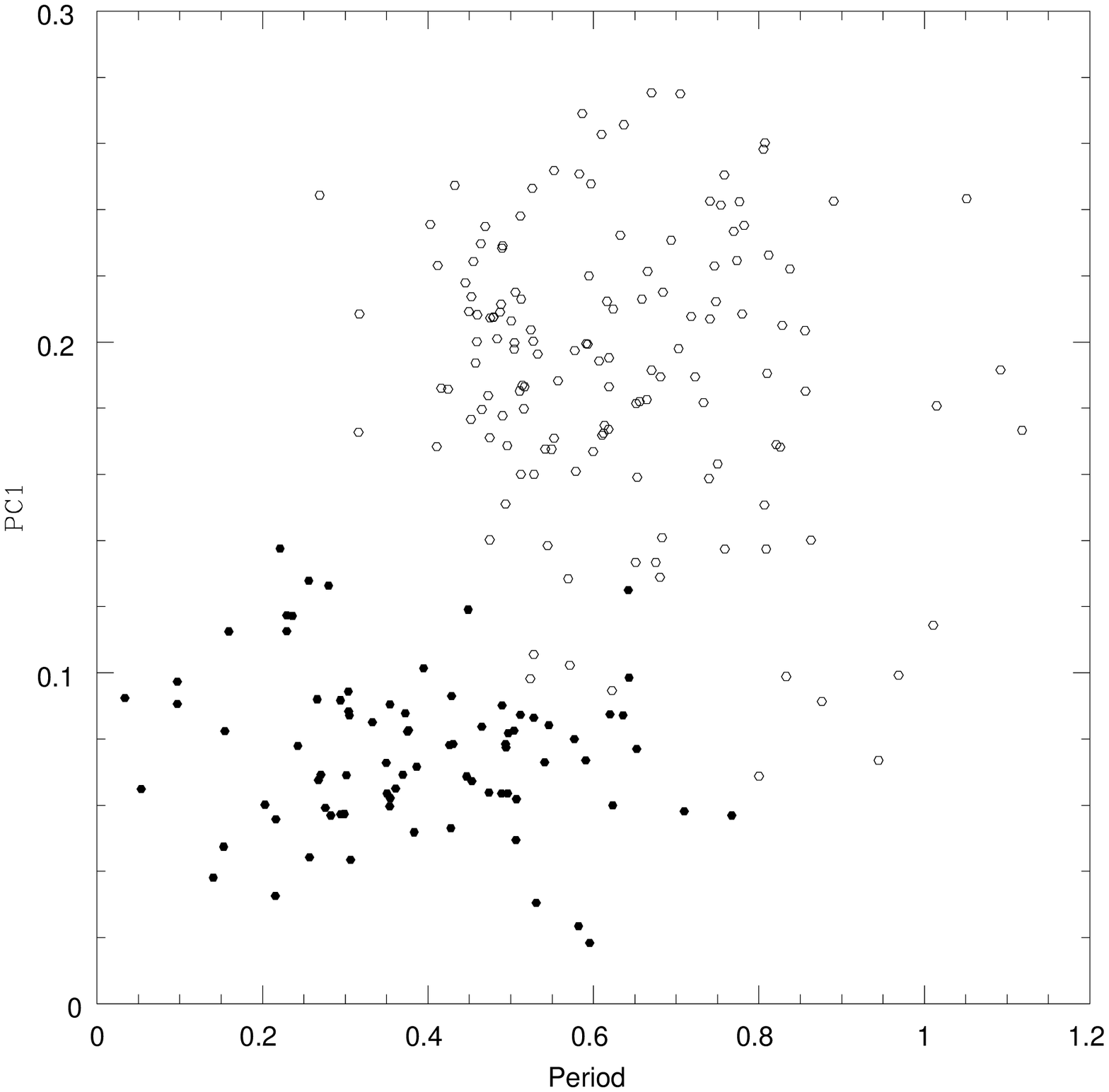,height=9cm,width=9cm}}
\caption{First Principal Component for LMC fundamental and first
overtone Cepheids from EROS}
\end{figure}

\begin{figure}
\centerline{\psfig{file=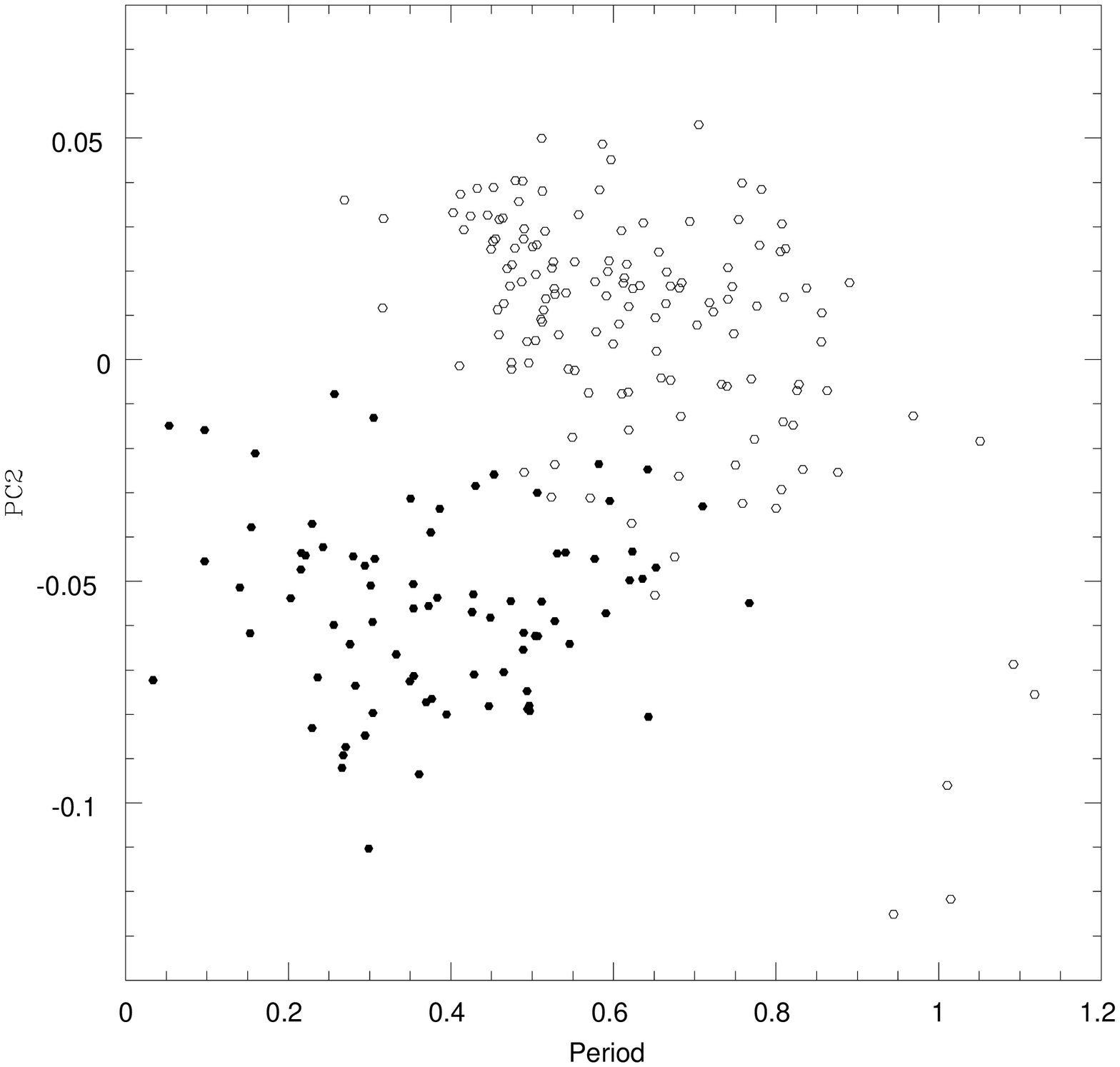,height=9cm,width=9cm}}
\caption{Second Principal Component for LMC
fundamental and first overtone Cepheids from EROS}
\end{figure}

\begin{figure}
\centerline{\psfig{file=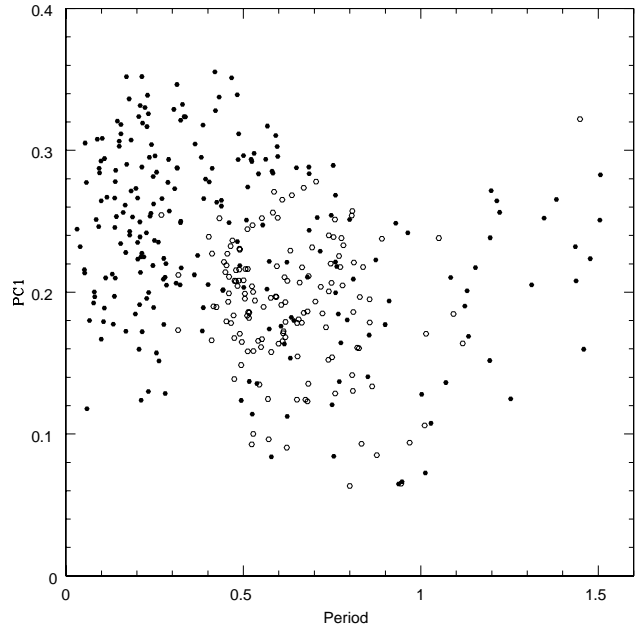,height=9cm,width=9cm}}
\caption{First Principal Component for LMC and SMC
fundamental mode Cepheids from EROS}
\end{figure}

\begin{figure}
\centerline{\psfig{file=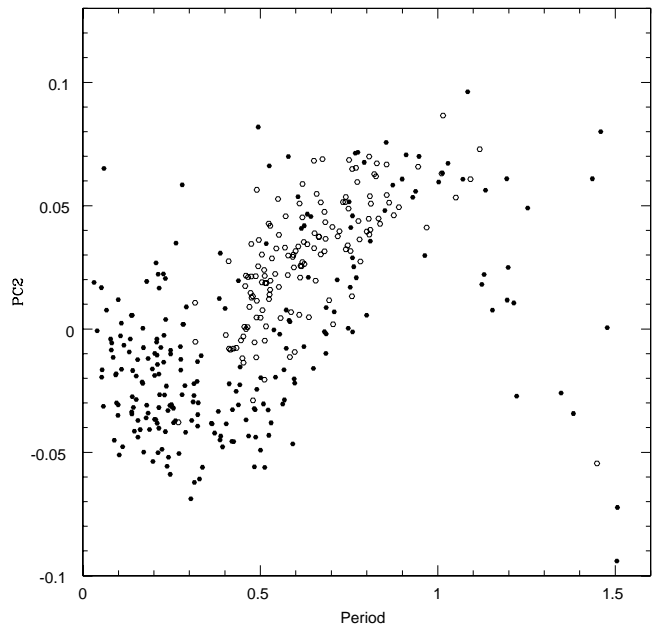,height=9cm,width=9cm}}
\caption{Second Principal Component for LMC and SMC
fundamental mode Cepheids from EROS}
\end{figure}

\begin{figure}
\centerline{\psfig{file=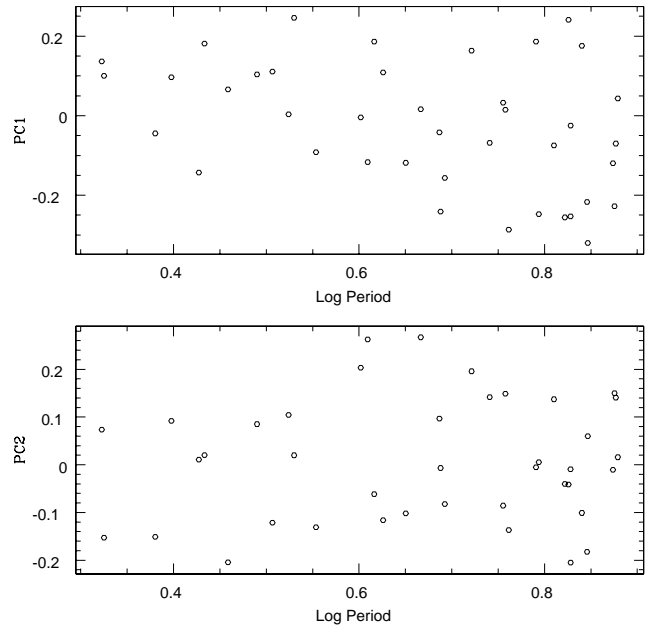,height=9cm,width=9cm}}
\caption{First two Principal Components for Galactic first overtone
Cepheids}
\end{figure}

\begin{figure}
\centerline{\psfig{file=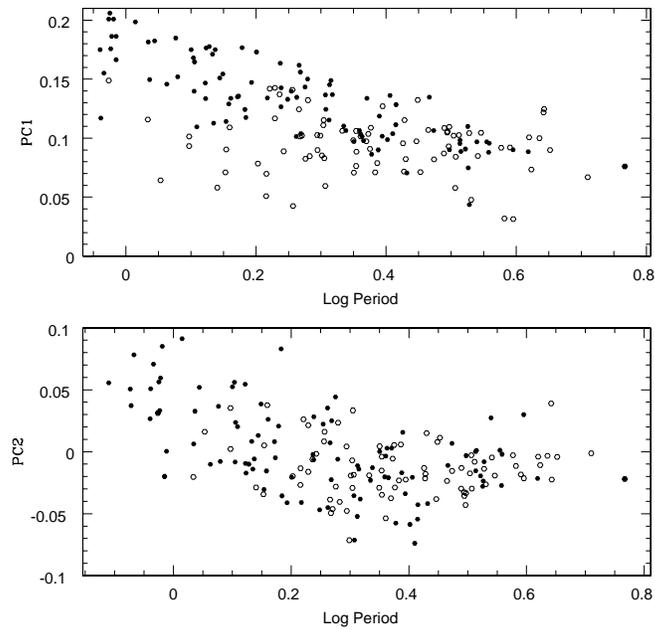,height=9cm,width=9cm}}
\caption{First two Principal Components for LMC and SMC first overtones.
Open and solid circles are the LMC and SMC respectively}
\end{figure}

\begin{figure}
\centerline{\psfig{file=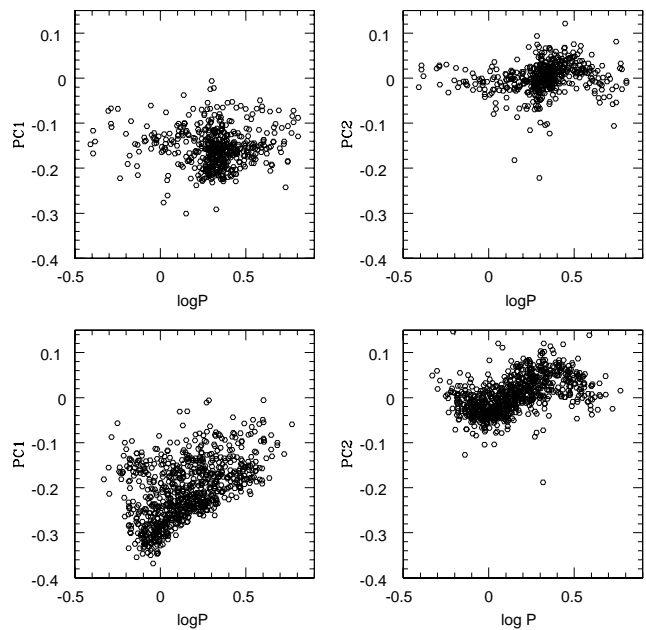,height=9cm,width=9cm}}
\caption{First two Principal Components for OGLE data for the
LMC (top panel) and SMC (bottom panel).}
\end{figure}

\end{document}